# Ultra-Conformable Free-Standing Capacitors Based on Ultrathin Polyvinyl Formal Films

Jonathan Barsotti*, Ikue Hirata, Francesca Pignatelli, Mario Caironi, Francesco Greco*, Virgilio Mattoli*

J. Barsotti, F. Pignatelli F. Greco, V. Mattoli
Center for Micro-BioRobotics @SSSA, Istituto Italiano di Tecnologia
Via R. Piaggio, 34, 56025 Pontedera, PI, (Italy)
E-mail: jonathan.barsotti@ymail.com; virgilio.mattoli@iit.it

I. Hirata, M. Caironi
Center for Nano Science and Technology @PoliMi, Istituto Italiano di Tecnologia
Via Pascoli 70/3, 20133 Milano, MI, (Italy)

F. Greco
Institute of Solid State Physics, Graz University of Technology, Petersgasse 16, 8010 Graz, Austria
E-mail: francesco.greco@tugraz.at

F. Greco
Department of Life Science and Medical Bioscience, Graduate School of Advanced Science and Engineering, Waseda University, 2-2 Wakamatsu-cho, Shinjuku-ku, 169-8480 Tokyo, Japan.

## Abstract -

Conformable Electronics refers to a class of electronic devices that have the ability to conformally adhere onto non-planar surfaces and materials, resulting particularly appealing for skin applications, such as the case of skin-worn unobtrusive (bio)sensors for healthcare monitoring. Conformability can be addressed by integrating basic electronic components on ultrathin polymeric film substrates. Among other basic electronic components, capacitors are fundamental ones for energy storage, sensing, frequency tuning, impedance adaptation and signal processing. In this work we present a novel approach for conformable capacitors based on a free-standing, ultrathin and ultra-conformable nanosheets of poly (vinyl formal) (PVF), which serve both as structural and dielectric component of the capacitor. A novel fabrication approach is proposed and applied to fully free-standing ultrathin capacitors fabrication, having an overall thickness as low as 200 nm; that represents, to the best of our knowledge, the thinnest free-standing capacitors ever reported. Thanks to the ultra-low thickness, the proposed capacitors are able to sustain flexure to extremely small curvature radii (as low as 1.5 µm) and to conform to complex surfaces, such as a nylon mesh with micrometric texture without compromising their operation.

## 1. Introduction

Organic electronics is enabling new technological applications in which electronic circuits are asked to be deformable, as in the case of flexible[1], stretchable[2] as well as conformable devices[3]. The great improvements in materials properties[4,5] and fabrication technologies[6–9] enabled constant developments toward applications, such as flexible displays[10], electronic paper, radio-frequency identification (RFID) tags, smart cards[11], electronic skin[12,13], wearable monitoring devices [14].

The term Conformable Electronics is used to refer to a class of electronic devices that, thanks to their reduced thickness and peculiar construction, conformally adhere onto non-planar surfaces and materials[15]. Thickness reduction also implies a total mass reduction and an increase in the aspect ratio values, factors that strongly affect adhesion and conformability. Under these conditions, Van der Waals forces become predominant, with a consequent adhesion improvement[16].

Conformability of devices is particularly appealing when skin is the target surface of interest. This is the case of skin-worn unobtrusive sensors, and more general of biosensors, to be used in biomedical, healthcare[15], sport activity and environment monitoring systems[15,17] applications. Indeed, in all cited applications it is highly demanded that devices are the least perceivable possible[18,19]. From a technological point of view empowering electronic devices with the ability to conformally adhere to a complex surface (such as skin) without compromising their functionality is very challenging. Mechanical instabilities such as buckling, crumpling, cracking, wrinkling, dewetting and swelling can set a severe limit to the deformation the device can sustain without losses of electrical performances, and eventually lead to complete failure. Moreover, since an obvious and main strategy to accomplish this task is to decrease the overall thickness and inherent stiffness of substrates and devices, there are intrinsic difficulties related to manipulation and processing of such thin systems. In order to try to overcome these difficulties, several interesting solutions have been proposed in literature.

In this vision, a novel temporary tattoo approach that makes use of commercial tattoo paper as a temporary substrate for ultra-conformable electrodes enabling an easy water-based transfer directly on skin, was proposed by Zucca *et al* [15]. In their work, the authors developed an unconventional way to fabricate ultra-conformable EMG electrodes using ultrathin conductive poly(3,4-ethyllendioxythiophene): poly(styrenesulfonate) (PEDOT:PSS) nanosheets. Following the temporary tattoo approach proposed by Zucca *et al.*, Ferrari *et al.* proposed a novel dry and perforable ultra-conformable skin-contact electrode[17]. Thanks to the tattoo-like nature, the unperceivable electrode can be directly transferred on cleaned skin as well as on hairy zones. Indeed, the ultralow electrode thickness allow hairs to perforate and grow through it, without significant performances degradation over a period as long as 48h. An analogous temporary tattoo approach has recently enabled novel applications in edible electronics[20] and organic photovoltaics (OPV)[21].

Similar applications have been enabled by using a spin coating or even a roll-to-roll (R2R) process for preparing free/standing PEDOT:PSS and PEDOT:PSS/ poly (D,L lactic acid) nanosheets[22–24].

Anyway, in order to address complex tasks, conformable sensors and devices, comprising Organic Light-Emitting Diodes (OLED), OPV and Organic Thin Film Transistors (OTFT)[25–27], must be integrated onto ultrathin polymer films. Nevertheless, basic electrical components such as resistances, inductances and capacitors are required as well. In particular, capacitors are fundamental components for energy storage, sensing, frequency tuning functionalities, impedance adaptation and signal processing. To this aim, a few groups tried to fabricate conformable capacitors.

A noteworthy example is provided by Luan *et al.*, who fabricated a free-standing epidermal super-capacitor assembling a $H_2SO_4$-PVA based electrolyte with two hybrid electrodes made by single-walled carbon nanotubes coated with PEDOT:PSS deposited by cyclic voltammetry[28]. Authors reported a good capacitance density value of 56 F $g^{-1}$, estimated considering the electrodes mass only, and demonstrated the possibility to bend and stretch the device, but they did not investigate conformability. Moreover, 1 µm was indicated as the lowest feasible thickness of these capacitors, since thinner films caused the occurrence of short circuit. Thickness (*t)* is indeed a key parameter in capacitor's design, which also strongly affects conformability. Maintaining suitable electrical properties when reducing the thickness of capacitors to very low values, such as sub-µm range is challenging.

In this work we propose a novel approach for fabricating conformable capacitors based on water-mediated (or air-mediated) lamination of free-standing ultrathin (UT) and ultra-conformable (UC) membranes, ensuring appealing impedance characteristics and spectral response. The capacitor is fabricated using free-standing, UT and UC nanosheets of poly (vinyl formal) (PVF), which serves both as structural (mechanical self-support, conformal adhesion) and as functional (dielectric) component of the capacitor. This approach enables a drastic reduction of the dielectric membrane thickness, down to 10 nm (170 nm of total thickness), and of mass, achieving a remarkable surface density of 1.48 mg/cm$^2$, while retaining a large surface area. With the same approach we demonstrated here the thinnest free free-standing UT capacitors ever reported, having an overall thickness (including top and bottom electrodes) as low as 200 nm, also showing very nice electrical performances on relatively large area (centimeter square scale). Moreover, capacitors show a wide working frequency, displaying a constant capacitance up to 280 kHz, with very low leakage current in the order of 10$^{-8}$ A/cm$^2$ at 0.5 MV/cm, and with an average dielectric strength of 1 MV/cm. The broad operative frequency bandwidth, combined with a high breakdown voltage, make these devices suitable for conformable electronic applications in which higher frequencies and voltages are required. In fact, they can provide an alternative to conformable electrolytic capacitors, typically having much lower cut-off frequencies and operation voltage limits[28,29]. Finally, thanks to the ultra-low thickness, the capacitors are able to sustain flexure to extremely small curvature radii (as low as 1.5 µm) and to conform on complex surfaces[30], such as a nylon mesh with micrometric texture.

Overall this work illustrates the potentialities of a nanosheet based approach in conformable capacitors fabrication, envisaging it as a powerful tool in realization of other truly conformable devices.

## 2. Results and Discussions

### *2.1. Capacitor structure*

UT capacitors have been developed by performing multiple recollections of various PVF nanosheet layers. To produce freestanding PVF nanosheets we used the spontaneous delamination process developed by Baxamusa *et al.*[31]. We selected PVF as host material because it is has very good dielectric properties, it is already adopted at industrial level (e.g. as insulator for electrical wires), and because Baxamusa *et al.* demonstrated that it is suitable to obtain robust ultrathin freestanding films.

Si wafer functionalization, implemented by spin coating deposition of a sub-nm layer of a cationic polyelectrolyte, Poly(diallyldimethylammonium chloride) (PDAC), enabled to deposit by spin coating PVF films that spontaneously delaminate and float when immersed into a water bath. A challenge we had to face in order to adopt such nanosheets to fabricate a capacitor was the deposition of the electrodes, since PVF nanosheets could sporadically display microscopic pin-holes, in which a metal electrode, such as sputtered Au, can easily penetrate, thus short-circuiting the device. Our strategy to avoid short circuit across electrodes was to pile up, through multiple recollection, three different layers for assembling a capacitor: a first PVF nanosheet carrying the bottom sputtered Au electrode (layer 1); an intermediate pure PVF dielectric membrane (layer 2); a top PVF nanosheet carrying the top Au electrode (layer 3). The full UT and UC capacitor fabrication process is shown in Figure 1 (a-c). By adopting this strategy, the probability of observing a short-circuit is drastically reduced, since it is very unlikely for two pin-holes of different layers to overlap when two nanosheets are recollected one on top of the other (see also Figure S1 in Supporting Information (SI) for PVF nanosheets pin-holes analysis).

Electrodes (layers 1 and 3) were fabricated by Au sputtering deposition ($t_{Au}$ = 40 nm) on a thin PVF nanosheet (nominal thickness $t_{sup}$ = 40 nm), before the release from the temporary substrate (total electrode thickness $t_1$, $t_3$ = 80 nm). A 40 nm-thick PVF nanosheet was found to be the thinnest one able to well support the Au sputtered electrode without being damaged by recollection/assembling process. A PVF layer ( $10 \leq t_2 \leq 160$ nm) composed the dielectric layer 2. After release in water, each nanosheet could be recollected onto solid surfaces (i.e. glass) or on plastic frames (i.e. PMMA rings) for further processing or assembling (see also Figure S3 in SI for floating nanosheets recollection sequence). After each recollection, all the nanosheets were dried in air in vertical position to facilitate water removal and under ambient conditions before recollection of subsequent layers. Performing multiple recollections of the various nanosheet layers, supported capacitors on glass were fabricated (Figure 1 (b)) in order to firstly asses devices functionality. Then, following the same recollection procedure, fully freestanding capacitors where assembled on plastic ring (Figure 1 (c)). The capacitor active area is determined by the orthogonal overlapping of the two electrodes forming a square with a nominal area of 25 mm$^2$. This value was corrected taking into account the actual covered Au surface, estimated by SEM images analysis (Figure S2 in SI). In fact, homogeneously distributed cracks were found in the Au deposited electrodes, reducing the actual conducting surface.

Pictures of supported and free-standing devices are shown in Figure 1 (d) and (e), respectively. Looking at the devices reported in Figure 1 (d), the PVF dielectric layer thickness variation can be appreciated thanks to the chromatic change due to light interference. The extreme colour uniformity in each sample is also a clear indication of the high uniformity of thickness of the dielectric nanosheets. The whole procedure is described in more detail in the Experimental Section (see also Figures S4 and S5 for more details on device structure).

## 2.2. Capacitors electrical characterization

Capacitance values will be reported in the following with the notation $C_x^t(f)$, where $x$ subscript indicates supported or free-standing ($x$ = *sup* or *free*, respectively) capacitor, $t$ is the dielectric (layer 2) thickness expressed in nm and $f$ is the frequency. The same notation will be used for cut-off frequency, $f_x^t$, defined as the frequency value at which capacitance reduces to the 80% of its value at 1 kHz, i.e. $C_x^t(f_x^t) = 0.8 \times C_x^t(1 \text{ kHz})$.

In Figure 2 (a) we report the capacitance at 1 kHz versus capacitor dielectric thickness in the range $10 \leq t_2 \leq 160$ nm. Data can be simply fit by adopting a parallel plates capacitor model, thus considering the capacity $C$ as follows:

$$C = \varepsilon_0 \varepsilon_{PVF} \frac{A}{t+h}, \tag{1}$$

where $\varepsilon_0$ is the vacuum dielectric constant, $\varepsilon_{PVF}$ is the PVF relative dielectric constant, $A$ is the capacitor area and $t$ is the thickness of dielectric (layer 2). The additional parameter $h$ describes the presence of the supporting PVF layer of the top electrode that acts as an extra dielectric layer, actually increasing the dielectric thickness. Thanks to this extremely low thickness, capacitors' aspect ratio spans from $10^{-5}$ to $10^{-4}$, fulfilling the flat capacitor model requirements.

The fit curve (red line, Figure 2 (a)) was obtained fixing $A$ and $h$ to the measured values, $A_m$ = 21.25 mm² and $h_m$= 41 ± 2 nm, respectively. The fit, performed using Eq. (1), allows to extract a dielectric constant value for $\varepsilon_{PVF}$ of 3.76 ± 0.08, with a determination coefficient value of $R^2$ = 0.9817. Such a determination coefficient value confirm the goodness of the fit. The calculated value of $\varepsilon_{PVF}$ is in perfect agreement with the value of 3.7 found by Khare *et al.*[32] and with values reported for various commercially available PVF materials[33].

In Figure 2 (b) a typical impedance module whit relative phase spectra for a capacitor with the thinnest PVF dielectric layer, $t_2$ = 10 nm, is shown. This is the thinnest PVF nanofilm we are able to produce, by releasing it in water and recollecting it on substrate while maintaining the film integrity on a centimeter scale. With such dielectric thickness we measured an average low frequency capacitance of 12.3 ± 0.6 nF at 1 kHz, corresponding to 49 ± 3 nF/cm². From 20 Hz to the cut-off frequency, falling at $f_{sup}^{10}$ = 120 ± 40 kHz as average on six devices, the Z module linearly decreases with slope -1, and the phase remains approximately constant at the value of 89. For $f > f_c$, the impedance of the device with $t_2$ = 10 nm tends to the series resistance ($R_s$ = 64 Ω) (originated by electrode and contact resistance) and with the phase correspondingly dropping down toward 0° (see also Figures S6 (a – h) for other impedance module and phase spectra). The corresponding capacitance spectrum of the capacitor with $t_2$ = 10 nm, which is flat until the cut-off frequency, is shown in the inset of Figure 2 (b). An operating range up to about 100 kHz is in the region of interest of many electronic applications, such as conformable sensors[34,35] and biosensors[36]. Indeed it is typically used in transduction of slowly varying physical and chemical quantities, typically lacking high frequency components.

Despite all supported devices were manually assembled, the shape of the capacitance spectra of the supported capacitors having different dielectric thickness $t_2$ showed very similar frequency dependence respect to the case with $t_2$ = 10 nm (see Figures S7 (a - h) in the SI), showing $f_c$ decreasing when $t$ increases (Figure 2 (c)), being cut-off frequency the inverse of the *RC* time constant of the devices. As an example, also the typical impedance module and phase spectra for a capacitor with the thickest PVF dielectric layer, ($t_2$ = 160 nm) is also reported in Figure 2 (d).

### 2.3. Leakage current and dielectric strength

Leakage current density $J$ of supported capacitors having different $t_2$ was measured and reported as a function of the applied electric field $E$ in Figure 2 (e). $J$ is plotted versus the effective PVF thickness $t_{eff} = t_2 + t_{sup}$, where $t_{sup}$ = 48 nm is the PVF electrode-supporting nanosheet thickness for the specific samples run. Tested devices showed comparable and remarkably low current densities in the range of $1 - 3 \times 10^{-8}$ A/cm$^2$ for $E \leq 0.5$ MV/cm. As an exception, the thickest device ($t_{eff}$ = 217 nm) showed even lower leakage current density, in the range of $6 \times 10^{-9} < J < 8 \times 10^{-9}$ A/cm$^2$ in the same conditions. The highest $J$ before device breakdown was observed in the thinnest device ($t_{eff}$ = 63 nm) and it was $5.7 \times 10^{-8}$ A/cm$^2$ at a remarkable $E$ = 0.84 MV/cm.

Devices' dielectric strength was tested by applying a linear positive voltage ramp until electrical breakdown. The breakdown points of the capacitors were clearly visible in the sudden increase of the current density for each curve. We obtained a remarkable average dielectric strength up to $E$ = 1.0 ± 0.3 MV/cm. Such result well compares with other commercially available PVF formulations such as Vinylec Series, with dielectric strength in the range of 0.26 – 0.39 MV/cm[33] and Formex, with a reported dielectric strength of 1.57 MV/cm[37]. This fact indicates that, thanks to our device fabrication strategy, the intrinsic PVF dielectric strength is not reduced in ultra-thin films, thus underlining the substantial absence of severe structural defects (see also Figure S8 in SI).

### 2.4. Free-standing capacitors

Free-standing capacitors were assembled by recollecting the various nanosheet layers as suspended membranes across plastic frames (Figure 1 (c)).With this approach we successfully produced freestanding capacitor with a $t_2$ down to 40 nm, thus with a total thickness of about 200 nm (see Figure S9 in SI for details). Typical impedance module and phase spectra for a free-standing capacitor with $t_2$ = 160 nm are shown in Figure 2 (f). As observed for supported devices, Z module linearly decreases with slope -1 and phase remains approximately constant to the value of 89° until $f_c$. Beyond cut-off the impedance tends to series resistance ($R_s$ = 23 Ω) and phase drops down toward 0°. The corresponding capacitance spectrum is reported in the inset of Figure 2 (f). For the free-standing device we measured an average capacitance $C_{free}^{160}$ = 4.32 ± 0.01 nF, corresponding to 17.28 ± 0.01 nF/cm² and a capacitance density of 0.4 mF·g⁻¹, computed on the whole device mass. The measure is in agreement with the value found for supported devices of the same thickness. Moreover, also free-standing capacitors show a flat capacitance curve from low frequencies until the cut-off point, consistently to what observed in supported devices. For the free-standing capacitors $f_c$ is generally higher than for the supported one at the same thickness. Indeed, the free-standing capacitors had an average cut-off frequency of $f_{free}^{160}$ = 1.43 ± 0.07 MHz, approximately one order of magnitude higher compared with supported devices ($f_{sup}^{160}$ = 0.28 ± 0.03 MHz). This difference is ascribable to the lower series resistance observed in the free-standing device probably due to contact differences originated by the device contacting (see Figure S10 in SI).

## 2.5. Mechanical test (SIEBIMM)

Mechanical properties of PVF nanosheets were evaluated by the "Strain-induced buckling instabilities for mechanical measurements" (SIEBIMM)[38]. This method allows measuring the Young's modulus of UT nanosheets and it has been applied to several materials [39,40]. An elastomeric substrate (PDMS) was pre-stretched and the nanosheet was recollected on it. By relaxing the pre-imposed strain, a quasi-periodic pattern of wrinkles was formed on the surface of the PVF nanosheet, aligned along a direction perpendicular to the applied pre-strain. By measuring the wavelength $\lambda$ of the wrinkles, it was possible to calculate the Young's modulus $E$ of the nanosheet from the equation

$$E_{PVF} = 3E_s \frac{1-v_{PVF}^2}{1-v_s^2} \left(\frac{\lambda}{2\pi t}\right)^3, \tag{2}$$

where the subscript $s$ stands for "substrate" (PDMS), $v$ is the Poisson's ratio, and $t$ is the nanosheet thickness.

In Equation (2) we used $E_s$ = 2.06 MPa (extrapolated from data reported in [41]) and $v_s$ = 0.5 for the PDMS substrate[41]. For PVF Poisson's ratio we used $v_{PVF}$ = 0.33, a value typically used for viscoelastic polymers (such as PVF)[42,43].

We performed SIEBIMM test on a PVF nanosheet with $t_2$ = 140 ± 2 nm. Surface profile measurements were carried out on wrinkled samples (pre-strain released) with an optical profilometer (DCM 3D, Leica) and analyzed by evaluating the Power Spectrum Density Function (PSDF), as described in SI (see Figure S11). From the PSDF analysis, we found a single main peak at k = 0.76 ± 0.06 $\mu m^{-1}$, corresponding to a wavelength $\lambda$ = 8.3 ± 0.7 $\mu$m. This value of $\lambda$ was used in Equation (2) obtaining a Young modulus for the PVF nanosheet ($E_{PVF}$) of 6 ± 2 GPa. Wang et al. found a Young's modulus value of 3.08 GPa for a membrane of poly(vinyl formal-acetal) as thick as 32 $\mu m$[44]. In addition, apparent Young's modulus values in the range 2.7 - 3.1 GPa are reported for a specific commercial PVF formulation (Formvar)[45]. Since there is no prior reference in literature to PVF nanosheets and to similar SIEBIMM measurements, a direct comparison with our results is not possible. However, the value we obtained is higher compared to cited data. Such an increase in $E$ could be explained by residual stress in the samples, easier to be induced because of the very low thickness. At the same time, we cannot *a priori* exclude that some relaxation effect of the polymeric chains, induced by the nanostructure, could take place affecting the mechanical properties of PVF nanosheets. Indeed behavior of nanoscale polymer thin films can significantly deviate from that of bulk materials[46]. While an accurate investigation of this phenomenon is beyond the scope of this work, it could help in future to clarify whether PVF shows such "confinement" effects.

## 2.6. Capacitors transferred onto target surfaces

In order to prove the transferability of our devices and the conformal adhesion to complex surfaces we transferred a free-standing capacitor sample with $t_2$ = 90 nm onto a nylon mesh (see optical image on Figure 3 (a) and SEM image of active area on Figure 3 (b)). It can be noticed that there is no significant difference in impedance module and phase spectra (see Figure 3(c)), as well as in the capacitance (inset), if compared with the general trends observed in both free-standing and supported devices. The capacitance of the transferred device is $C_{free}^{90}$ = 5.9 nF, in agreement with the value of $C_{sup}^{90}$ = 6.2 nF measured for supported device with same thickness. The cut-off frequency we measured for the transferred capacitor ($f_{mesh}^{90}$) is 0.36± 0.01 MHz, higher than the one observed for supported device with similar thickness. A SEM image of a bare PVF nanosheet collected on the mesh and an image of the mesh (inset) are reported in Figure 3 (d) in order to show PVF conformability on such a complex surface. It is noteworthy how the Au electrode is affecting the conformability, by comparing micrographs of the electrode area (stacked PVF and double electrode layers, Figure 3 (c)) with the sole PVF layer (Figure 3 (d)). This demonstrates that the free-standing capacitor maintained its structural integrity and it was still perfectly working after the transfer onto a complex micro-structured surface such as the mesh's one.

In order to evaluate conformability characteristics of the device, a freestanding UT capacitor was transferred on an extremely sharp object, such as a scalpel (inset: Figure 3 (e)). By means of a Focused Ion Beam (FIB) milling we excavated a groove in the transferred device. In Figure 3 (e) a SEM image of the groove is reported. The image shows the device adaptation along the narrow scalpel cutting edge, where the curvature radius $r$ reaches its smaller value of approximately 1.5 µm, as graphically estimated by SEM image. Along the clean groove edge the device layered structure is revealed, not showing any damage in the layers despite local bending at $r \approx 1.5$ µm. The layer structure is visible with the darker PVF dielectric layer sandwiched between the two conductive and brighter Au electrodes. Top electrode had an additional Au layer, deposited after device transfer, for the sake of improving quality of SEM imaging. In order to compare capacitor structure in the bent case with the flat case of the supported one (on glass) a SEM image of a supported capacitor cross section is also reported in Figure 3 (f).

## 3. Conclusions

In this work we proposed a novel approach to fabricate UT and UC free-standing capacitors based on a novel nanosheet approach, where a PVF layer is employed as both dielectric and structural conformable layer. PVF was selected as a candidate material thanks to its unique combination of good electrical properties with excellent mechanical resistance. By using this approach, free-standing capacitors with various PVF dielectric layer thickness in the range from 10 nm to 160 nm, and Au-PVF electrodes were fabricated. Free-standing capacitors showed extremely good conformability, sustaining micrometric curvature radii. This capability was combined with remarkable mass and capacitance densities of 1.48 mg/cm$^2$ and 0.4 mF·g$^{-1}$ respectively, with RC limited cut-off frequency extending up to the MHz, leakage density currents as low as $6\times10^{-9}$ A/cm$^2$ and breakdown field of 1.0 MV/cm. To show actual devices capabilities a capacitor with a dielectric layer as thin as 90 nm was transferred on a micro-structured fabric mesh, as an example of surface with complex topography at the microscale. Device characterization performed after the transfer revealed a correct device operation, consistent with that observed on same devices supported on solid substrates. Once released, the capacitors maintained their electrical characteristics while conforming to the target surface, thus confirming the success of the proposed approach. Finally, by transferring a free-standing capacitor onto a scalpel blade we quantitatively demonstrated the extremely high conformability of the structure that is able to sustain a curvature radius as low as 1.5 µm without breaking. This candidates our UC free standing capacitors to be used as deformation sensor, storage energy system as well as to be integrated in more complicated conformable circuits, tracing a promising path towards a new generation of truly unperceivable and ultra-conformable electronic circuits and systems. Finally, it is worth to note that ultrathin freestanding dielectric lamination approach can be directly extended to build other devices requiring high quality dielectric, such as low voltage organic transistors, an application that is under investigation in our group.

# 4. Experimental Section

*4.1 Materials*

Poly(vinyl formal) (PVF) was purchased from Sigma-Aldrich(trade name Vinylec K ). Ethyl lactate (EL, ≥ 98%, FCC, FG) and polydiallyldimethylammonium chloride (PDAC, 20 wt% in water solution) were purchased from Sigma-Aldrich. Si wafer (3" diameter and are 381 ± 25 µm thick) used for film fabrication were purchased by Silicon Materials (Si-Mat). Supported capacitor were assembled on Plain Micro-Slides, used as glass substrate and purchased by VWR International. Poly(dimethylsiloxane) (PDMS, Sylgard 184 silicone elastomer base and curing agent) was purchased by Dow Corning Corp. The nylon mesh (Sefar Nytal HC-58) was purchased from Sefar Co. Ltd., Switzerland. The Ag paste used for contacting all the devices was a bi-component Conductive Silver Epoxy paste, purchased by CircuitWorks®.

*4.2 Methods*

**PVF nanosheets preparation.** A PDAC solution at 0.5 % wt was prepared by diluting the commercial 20 % wt water solution with deionized water. PDAC solution was filtered immediately before use with a cellulose acetate filter ( 0.45 µm pores diameter, Minisart[TM]). PVF solutions (1 %, 1.5 %, 2 %, 2.5 %, 3 %, 3.5 %, and 4 % in weight) were prepared by dissolving PVF in EL according to a procedure presented elsewhere[31]. Solutions were stirred at 650 rpm until all PVF was dissolved. PVF Solutions were filtered immediately before use, with a hydrophobic filter (0.2 µm pores diameter, Minisart[TM]).

A Si wafer was first functionalized with a subnanometric PDAC layer[31]. PDAC solution was spin coated at 4000 rpm for 15 s and then baked at 100 °C for 10 s on a hotplate. About 2 ml of PDAC solution were deposited for each nanosheet spinning session. Extra PDAC was rinsed with deionized water and then the surface was dried with a compressed air gun. PVF solution was then spin coated over the PDAC layer in a 2-steps process of a total duration of 10 s. During the first 5 s, spin coating speed was set at 300 rpm while during the remaining 5 s it was set at 3000 rpm. About 2 ml were deposited for each nanosheet PVF spinning session (see also Figure S11 in SI for PVF spinning characterization curve). PVF was then baked for 60 s at 50 °C on a hotplate in order to remove extra solvent. PVF nanosheets thickness characterization was performed by means of a P-6 stylus profilometer (KLA Tencor, USA) averaging over 15 estimated values randomly acquired on the film surface before the release from Si wafer. Same measurements were carried out on nanosheets after release in water and recollection onto a fresh Si wafer. Anyway, no differences were found in PVF nanosheet thickness before and after PVF nanosheet release/recollection.

***Capacitor assembly.*** We used plain micro-slide glass substrates for assembling supported capacitors and ring shaped polymethylmetacrylate circular frames (with 1 cm diameter of inner-hole) for recollecting and assembling the free-standing membrane capacitors. Both top and bottom electrodes were fabricated on 40 nm-thick PVF nanosheets supported on Si wafer, by Au sputtering deposition (70 W for 25 s at $10^{-2}$ mbar in Ar; nominal Au thickness $t_{Au}$ = 40 nm) through a physical mask (5 mm x 20 mm). A custom-made sputtering system assembled with Leybold components by Sistec was used. Rectangular electrode layers were shaped cutting the nanosheet with a surgical blade and then delaminated in water using a home-made immersing system for a controlled immersion of wafers at the fixed angle of 40°.

Supported capacitors were manually assembled by immersing the glass substrate into water. Recollection of the floating bottom electrode was performed by pulling the glass slide out of the water with a fixed angle (ca. 90°) and allowing one edge of floating electrode to adhere to the glass and to be consequently pulled away from water surface; see Figure S2 in SI. A thin water film between recollected nanosheet and glass substrate facilitated film relaxation. The PVF dielectric layer was then recollected on top of the bottom electrode (previously recollected and dried), following the same recollection procedure. Finally, top electrodes were recollected on top of the PVF dielectric layer with the same recollection procedure. Electrodes recollection was performed maintaining the right angle between electrode Au strips (approximately 90°) with an overlapping area of 25 mm$^2$. Before each successive recollection step, the previously recollected layer was dried under ambient conditions to avoid nanosheet delamination during the successive recollection. Free-standing capacitors were also manually assembled with a similar stepwise procedure on top of circular frames. Once all the layers of capacitors (both supported and free standing) were recollected, electrical contacts were made by fixing electrical wire to Au electrodes by means of conductive Ag paste (CircuitWorks® Conductive Silver Epoxy).

The percentage of the electrode area covered by Au was estimated to be around 85 % by analyzing the SEM images of electrode surfaces with MatLab software through thresholding operation (details reported in SI). An actual area of 25 mm$^2$ x 0.85 = 21.25 mm$^2$, was used in the calculations reported in the work.

***Capacitor electrical characterization.*** Impedance and capacitance spectra of on-glass supported capacitors were recorded in the frequency range 20 Hz - 2 MHz, by means of an E4980AL Precision LCR Meter (Agilent, now Keysight Technologies). Capacitors with 10 ≤ $t_2$ ≤ 160 nm were characterized. Capacitance spectra were calculated from impedance module and phase, describing the device with an equivalent parallel circuit between an ideal resistor and an ideal capacitor.

***Capacitors transfer onto mesh.*** The free-standing capacitor was transferred on the nylon mesh with a wet process in order to attain improved conformability to the mesh surface. We previously moistened the mesh with water. We fixed the mesh onto a rigid transparent substrate of PMMA to manipulate it and to electrically contact the device with conductive Ag paste.

***Mechanical test (SIEBIMM).*** Si wafers were functionalized in a bell desiccator with chlorotrimethylsilane vapours for 60 minutes in order to improve PDMS detachment. Crosslinking of PDMS was obtained by mixing the PDMS elastomer base with the curing agent (10:1 ratio by weight). The mixture was then degassed in a vacuum bell desiccator in order to remove air bubbles. PDMS was then deposited by spin coating onto a pre-functionalized Si wafer at a speed of 200 rpm for 60 s and cured at T = 95 °C for 60 min in a convection oven. The cured PDMS was cut into rectangular slabs (4 x 1.5 x 0.3 cm$^3$) and pre-stretched to 5%. PVF nanosheets were recollected onto pre-stretched PDMS slabs from water and then dried in vacuum at RT overnight prior to the SIEBIMM test. The strain of the PDMS substrate was finally relaxed, producing the buckling of the PVF nanosheet. The buckling wavelength of the nanosheet was measured by an optical profilometer (DCM 3D, Leica), operating in confocal mode with an objective EPI 150X-L. The formula used to calculate the Young's modulus of the PVF nanosheets is reported in equation (2).

## Acknowledgements

We acknowledge Mr. Carlo Filippeschi for Au sputtering deposition and FIB Dual Beam operations. M.C. acknowledges support by the European Research Council (ERC) under the European Union's Horizon 2020 research and innovation program "HEROIC," Grant Agreement No. 638059. F.G. acknowledges financial support from Top Global University Program at Waseda University, Tokyo from MEXT Japan. V.M. acknowledges support by the RoboCom++ FLAG-ERA JTC 2016 project.

## References

[1]  M. Caironi, Y.-Y. Noh, Eds. , *Large Area and Flexible Electronics*, Wiley-VCH, **2015**.

[2]  T. Sekitani, T. Someya, *Adv. Mater.* **2010**, *22*, 2228.

[3]  T. Someya, S. Bauer, M. Kaltenbrunner, *MRS Bull.* **2017**, *42*, 124.

[4]  J. Onorato, V. Pakhnyuk, C. K. Luscombe, *Polym. J.* **2016**, *49*, 1.

[5]  Y. S. Rim, S. H. Bae, H. Chen, N. De Marco, Y. Yang, *Adv. Mater.* **2016**, *28*, 4415.

[6]  M. Eslamian, *Nano-Micro Lett.* **2017**, *9*, 1.

[7]  D.-H. Kim, R. Ghaffari, N. Lu, J. A. Rogers, *Annu. Rev. Biomed. Eng.* **2012**, *14*, 113.

[8]  a. a. Bessonov, M. N. Kirikova, *Nanotechnologies Russ.* **2015**, *10*, 165.

[9]  J.-H. Ahn, J. H. Je, *J. Phys. D. Appl. Phys.* **2012**, *45*, 103001.

[10]  Z. Li, H. Meng, Eds. , *Organic Light-Emitting Materials and Devices*, CRC Press, **2007**.

[11]  F. M. Li, A. Nathan, Y. Wu, B. S. Ong, *Organic Thin Film Transistor Integration*, John Wiley & Sons, **2011**.

[12]  S. Bauer, *Nat. Mater.* **2013**, *12*, 871.

[13]  M. L. Hammock, A. Chortos, B. C. K. Tee, J. B. H. Tok, Z. Bao, *Adv. Mater.* **2013**, *25*, 5997.

[14]  Y. Liu, M. Pharr, G. A. Salvatore, *ACS Nano* **2017**, 11, 9614.

[15]  A. Zucca, C. Cipriani, Sudha, S. Tarantino, D. Ricci, V. Mattoli, F. Greco, *Adv. Healthc. Mater.* **2015**, *4*, 983.

[16]  B. N. Chapman, *J. Vac. Sci. Technol.* **1974**, *11*, 106.

[17]  L. M. Ferrari, S. Sudha, S. Tarantino, R. Esposti, F. Bolzoni, P. Cavallari, C. Cipriani, V. Mattoli, F. Greco, *Adv. Sci.* **2018**, *1700771*.

[18]  T. Sekitani, T. Someya, *Mater. Today* **2011**, *14*, 398.

[19]  M. Kaltenbrunner, T. Sekitani, J. Reeder, T. Yokota, K. Kuribara, T. Tokuhara, M. Drack, R. Schwödiauer, I. Graz, S. Bauer-Gogonea, S. Bauer, T. Someya, *Nature* **2013**, *499*, 458.

[20]  G. E. Bonacchini, C. Bossio, F. Greco, V. Mattoli, Y.-H. Kim, G. Lanzani, M. Caironi, *Adv. Mater.* **2018**, 1706091.

[21]  N. Piva, F. Greco, M. Garbugli, A. Iacchetti, V. Mattoli, M. Caironi, *Adv. Electron. Mater.* **2017**, *1700325*, 1.

[22]  F. Greco, A. Zucca, S. Taccola, A. Menciassi, T. Fujie, H. Haniuda, S. Takeoka, P. Dario, V. Mattoli, *Soft Matter* **2011**, *7*, 10642.

[23]  F. Greco, A. Zucca, S. Taccola, B. Mazzolai, V. Mattoli, *ACS Appl. Mater. Interfaces* **2013**, *5*, 9461.

[24]  A. Zucca, K. Yamagishi, T. Fujie, S. Takeoka, V. Mattoli, F. Greco, *J. Mater. Chem. C* **2015**, *3*, 6539.

[25]  M. S. White, M. Kaltenbrunner, E. D. Głowacki, K. Gutnichenko, G. Kettlgruber, I. Graz, S. Aazou, C. Ulbricht, D. A. M. Egbe, M. C. Miron, Z. Major, M. C. Scharber, T. Sekitani, T. Someya, S. Bauer, N. S. Sariciftci, *Nat. Photonics* **2013**, *7*, 811.

[26]  M. Kaltenbrunner, M. S. White, E. D. Głowacki, T. Sekitani, T. Someya, N. S. Sariciftci, S. Bauer, *Nat. Commun.* **2012**, *3*, 770.

[27]  T. Sekitani, U. Zschieschang, H. Klauk, T. Someya, *Nat. Mater.* **2010**, *9*, 1015.

[28]  P. Luan, N. Zhang, W. Zhou, Z. Niu, Q. Zhang, L. Cai, X. Zhang, F. Yang, Q. Fan, W. Zhou, Z. Xiao, X. Gu, H. Chen, K. Li, S. Xiao, Y. Wang, H. Liu, S. Xie, *Adv. Funct. Mater.* **2016**, 26, 8178.

[29]  R. Kötz, R. Kötz, M. Carlen, M. Carlen, *Electrochim. Acta* **2000**, *45*, 2483.

[30]  L. B. Freund, S. Suresh, *Thin Film Materials: Stress, Defect Formation and Surface Evolution*, Cambridge University Press, **2003**.

[31]  S. H. Baxamusa, M. Stadermann, C. Aracne-Ruddle, A. J. Nelson, M. Chea, S. Li, K. Youngblood, T. I. Suratwala, *Langmuir* **2014**, *30*, 5126.

[32]  P. K. Khare, R. S. Chandok, *Polym. Int.* **1995**, *38*, 153.

[33]  SPI Supplies, "Vinylec ® Resins data sheet,"

[34]  W. Lee, D. Kim, N. Matsuhisa, M. Nagase, M. Sekino, G. G. Malliaras, T. Yokota, T. Someya, *Proc. Natl. Acad. Sci.* **2017**, 114, 10554.

[35]  T. Yokota, Y. Inoue, Y. Terakawa, J. Reeder, M. Kaltenbrunner, T. Ware, K. Yang, K. Mabuchi, T. Murakawa, M. Sekino, W. Voit, T. Sekitani, T. Someya, *Proc. Natl. Acad. Sci.* **2015**, *112*, 14533.

[36]  A. M. Pappa, O. Parlak, G. Scheiblin, P. Mailley, A. Salleo, R. M. Owens, *Trends Biotechnol.* **2018**, *36*, 45.

[37]  M. M. Sprung, F. O. Guenther, M. T. Gladstone, *Ind. Eng. Chem.* **1955**, *47*, 305.

[38]  C. M. Stafford, C. Harrison, K. L. Beers, A. Karim, E. J. Amis, M. R. VanLandingham, H. C. Kim, W. Volksen, R. D. Miller, E. E. Simonyi, *Nat. Mater.* **2004**, *3*, 545.

[39]  S. Taccola, V. Pensabene, T. Fujie, S. Takeoka, N. M. Pugno, V. Mattoli, *Biomed. Microdevices* **2017**, *19*, 51.

[40]  F. Greco, A. Bellacicca, M. Gemmi, V. Cappello, V. Mattoli, P. Milani, *ACS Appl. Mater. Interfaces* **2015**, *7*, 7060.

[41]  I. D. Johnston, D. K. McCluskey, C. K. L. Tan, M. C. Tracey, *J. Micromechanics Microengineering* **2014**, *24*, 35017.

[42]  G. N. Greaves, A. L. Greer, R. S. Lakes, T. Rouxel, *Nat. Mater.* **2011**, *10*, 986.

[43]   N. W. Tschoegl, W. G. Knauss, I. Emri, *Mech. Time-Dependent Mater.* **2002**, *6*, 3.

[44]   Q. Wang, Y. Yin, H. Xie, J. Liu, W. Yang, P. Chen, Q. Zhang, *Soft Matter* **2011**, *7*, 2888.

[45]   A. A. Tracton, *Coatings Technology Handbook*, **2006**.

[46]   J. L. Keddie, R. A. L. Jones, R. A. Cory, *Faraday Discuss.* **1994**, *98*, 219.

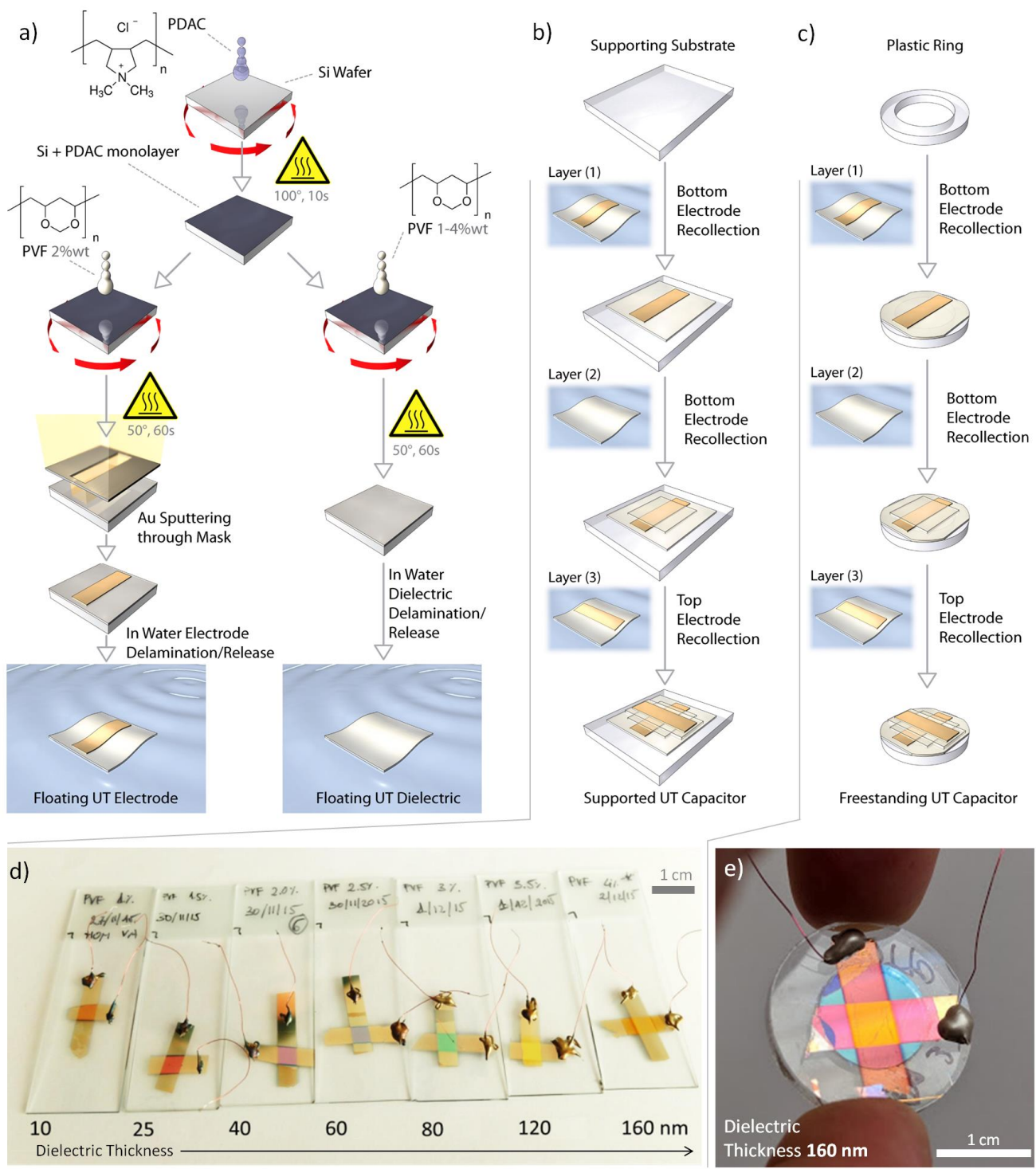

**Figure 1.** a) Scheme of PVF nanosheet fabrication process and release in water. Left branch refers to bilayer Au-PVF ultrathin electrode (layers 1 and 3), while right branch refers to pristine PVF ultrathin dielectric (layer 2) fabrication. Scheme of assembly by consecutive recollection of free-standing floating nanosheets: b) ultrathin supported capacitor on glass substrate and c) free-standing capacitor membrane on a plastic ring frame. d) UT supported capacitor samples, ordered by increasing dielectric layer thickness $t_2$ (structural colors due to thin film visible light interference). e) UT free-standing capacitor membrane on a plastic ring frame.

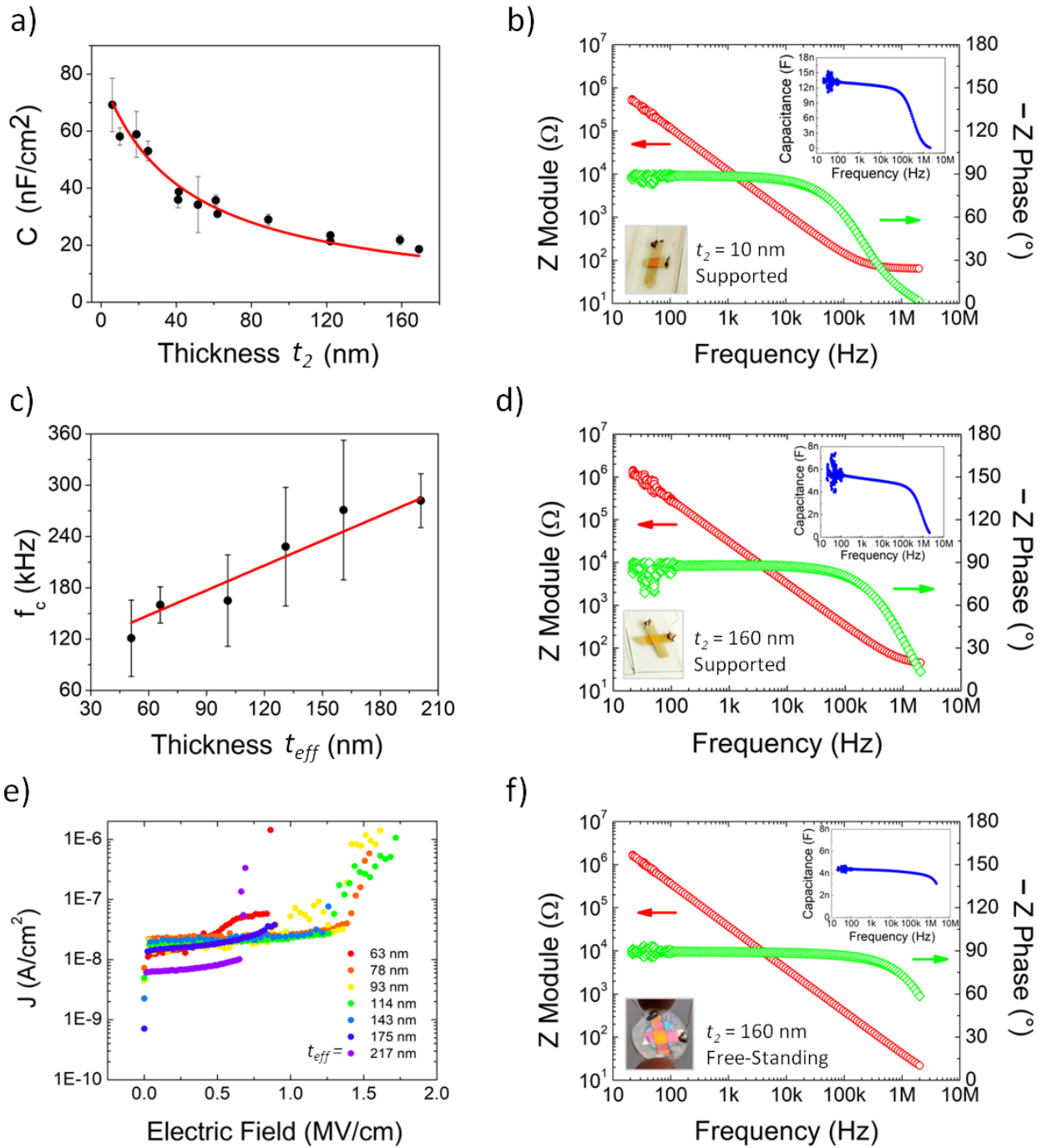

**Figure 2.** a) Capacitance at 1 kHz C of supported capacitors as a function of PVF dielectric thickness $t_2$ (layer 2). Data fit (red line) was computed using flat capacitor model and returned a relative dielectric constant value of ε = 3.76. c) Cut-off frequency $f_c$ as a function of $t_{eff}$ ($t_2 + t_{sup}$); fit curve (red line) highlights the linear dependence. e) Leakage current density $J$ as a function of the applied electric field $E$ for supported capacitors with different $t_{eff}$; device breakdown occurs in correspondence of sudden current increase in each device curve. Typical examples of impedance $Z$ module and phase spectra of: b) supported capacitor with $t_2$ = 10 nm; d) supported capacitor with $t_2$ = 160 nm; f) free-standing capacitor with $t_2$ = 160 nm (Insets: corresponding capacitance spectra computed applying series capacitor-resistor model (right), and optical image(left)).

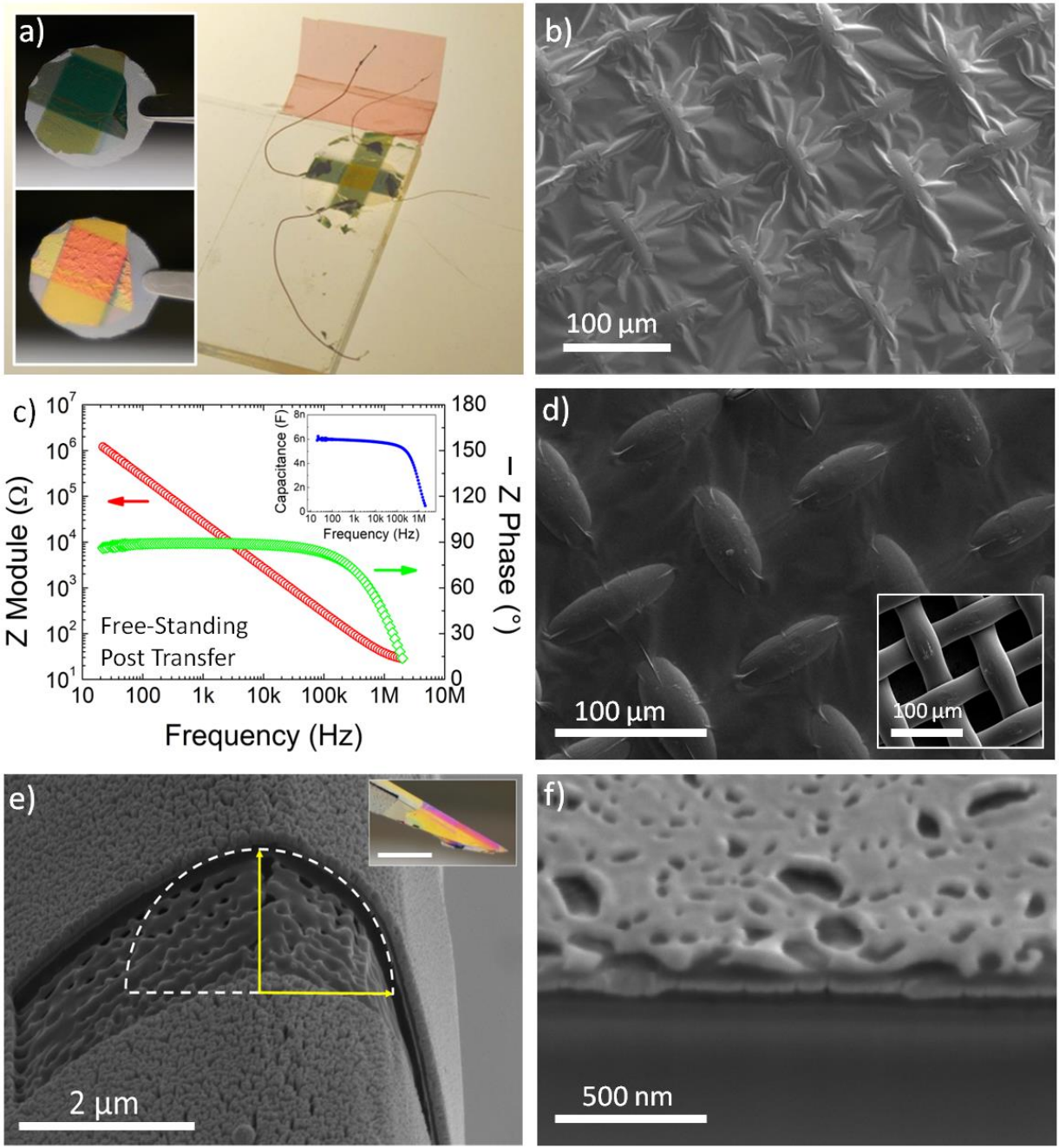

**Figure 3.** Optical image (a) and SEM image (b) of a free-standing capacitor (PVF dielectric $t_2$ = 90 nm) transferred onto a nylon mesh. The mesh was fixed to a plastic substrate for handling (Insets in (a): optical images the same capacitor at different illumination angles. c) Impedance module and phase spectra of the capacitor transferred onto the nylon mesh (Inset: corresponding capacitance spectrum computed applying series capacitor-resistor model, top-right corner). d) SEM image of a bare PVF nanosheet ($t_2$ = 90 nm) conformally adhering onto the mesh pattern (inset: SEM image of the bare mesh). e) SEM image of a capacitor ($t_2$ = 60 nm) recollected on a scalpel (Inset: optical image of the conformed free-standing capacitor on the scalpel, scale bar 4 mm). The image shows a trench obtained by Focused Ion Beam (FIB) milling of the sample surface in correspondence of the scalpel blade; the trench reveals the local curvature radius (white dashed semicircle) sustained by the device, of about 1.5 µm (yellow arrows). f) SEM image of a cross-section obtained by FIB milling on a capacitor supported on glass ($t_2$ = 90 nm).